# Dependence of Magnetic Anisotropy and Magnetoresistance of Ni$_{81}$Fe$_{19}$-Films on Annealing

Tilmann Lorenz, Andrea Käufler, Yuansu Luo, Michael Moske and Konrad Samwer

*Institut für Physik, Universität Augsburg, D-86135 Augsburg, Germany*

**ABSTRACT**

Permalloy (Py:Ni$_{81}$Fe$_{19}$) exhibits an anisotropic magnetoresistance (AMR) which is very often used to read magnetic signals from storage devices. Py-films of thickness 20nm were prepared by dc-magnetron sputtering in a magnetic field onto thermally oxidized Si-wafers and annealed ex situ at temperatures up to 1000K in order to investigate the dependence of the magnetic anisotropy and the AMR on heat treatments. The films exhibit an uniaxial anisotropy after preparation which changes during annealing above 520K. The AMR along the former magnetically easy axis as well as the corresponding field sensitivity are increased by a heat treatment around 700K reaching maxima of about 8% and a maximum sensitivity of 1.5%/Oe, respectively. We discuss possible sources for the change in anisotropy, i.e. strain effects, inhomogeneities, and changes of the local atomic order.

PACS: 75.30.Gw, 75.50.Bb, 75.70.-i, 81.40.Rs

Keywords: Magnetic anisotropy, magnetic films, permalloy, magnetoresistance

## I. INTRODUCTION

Known for more than 100 years [1] the anisotropic magnetoresistance (AMR) seemed to be the optimal method to write and read magnetic signals for storage devices. Because of its soft magnetic properties Permalloy (Py:$Ni_{81}Fe_{19}$) was one of the most protruding materials for this purpose. Even though magnetic multilayers [2], granular systems [3], spin valve systems [4] and metal-insulator transition systems [5] nowadays show much higher changes in resistivity by applying a magnetic field it was only shown recently that spin valves [6], discontinuous Py/Ag-multilayers [7] and conventional Py/Au-multilayers [8] could reach field sensitivities of about 1%/Oe which are even higher than the sensitivities obtained from AMR of "pure" Py. Here we discuss the effect of heat treatments on pure Py-films prepared in a magnetic field. It is known [9] that Py annealed at temperatures of 520K and above may exhibit very complicated magnetic properties, especially a change of the anisotropy. This is at about the same annealing temperature at which the magnetoresistive behavior of the discontinuous multilayers starts to change [7,10].

## II. EXPERIMENTAL DETAILS

Py-films of thickness 20nm were dc-magnetron sputtered at ambient temperatures onto thermally oxidized Si-wafers. The background pressure and the Ar pressure during preparation were $1 \times 10^{-8}$mbar and $9 \times 10^{-3}$mbar, respectively, the deposition rate was 3.0nm/s. The total thickness was controlled ex situ by small angle X-ray diffraction measurements. Because the distance between target and substrate is only 5 cm, the films are sputtered within the stray field of the Py-target and the magnets of the sputter gun. In this distance we measure a maximum static magnetic field of $15 \times 10^{-4}$T using a Hall device. Because of the rotational symmetry of our sputtering device and one single rotation of the substrate underneath the target during the entire film preparation, this field changes its direction relative to the sample position during the sputter process and therefore with film thickness. After preparation the wafers are cut into pieces of 1x1cm$^2$ and the samples are annealed at pressures below 5x10$^{-6}$mbar for 15 minutes up to 1000K. The oven exhibits an ac-magnetic field of about 50x10$^{-4}$T during heating. By comparing the results of samples annealed within this ac-field along different directions we found no direction dependence on this field. For the investigation of the deposition field influence on the anisotropy we did not apply a dc-field during the anneals.

The resistivity was measured in a four probe geometry (with spring loaded contacts at the corners of the samples) in fields up to 0.1 Tesla. In order to investigate the AMR as well as the inplane anisotropy, we measured each sample in four configurations as shown in Fig. 1, i.e. with the sensing current I as well as the orientation of the easy and the hard axis parallel and perpendicular to the external field H and to each other. All measurements were performed only at room temperature with the magnetic field always inplane.

The AMR is usually calculated by $(\rho_s^{\parallel} - \rho_s^{\perp})/\rho_{av}$ with $\rho_s^{\parallel}$ and $\rho_s^{\perp}$ as the resistivity at magnetical saturation ( I parallel "$\parallel$" and perpendicular "$\perp$" to the external field), normalized by $\rho_{av}$ = 1/3 $\rho_s^{\parallel}$ + 2/3 $\rho_s^{\perp}$ to represent the condition of zero magnetization [11]. However, in our measurements there is an uncertainty of the absolute resisitivity values due to an uncertainty of the electrical contact positions on the sample. Therefore we calculate the AMR from the normalized maximum resisitivity change for each individual measurement:

$$\left(\frac{\Delta\rho}{\rho}\right)_{AMR} = \frac{1}{2}\left[\left(\frac{\rho_s^{\parallel} - \rho_{min}^{\parallel}}{\rho_{min}^{\parallel}}\right) + \left(\frac{\rho_{max}^{\perp} - \rho_s^{\perp}}{\rho_{max}^{\perp}}\right)\right] \quad (1)$$

This definition does not produce higher values than the commonly known one. As the best experimentally accessible values for $\rho_{min}^{\parallel}$ and $\rho_{max}^{\perp}$ we take the resistivity $\rho_c$ at the coercitive field which should be closest to these values. Furthermore, we define the field sensitivity $(\Delta\rho/\rho_c)$/FWHM as the normalized resistivity change divided by the full width at half maximum (FWHM) of the respective resistivity measurement as it is also used by other authors [7].

For further magnetic characterization we also performed magnetization measurements in a VSM (EG&G Princeton Applied Research) at room temperature.

**III. RESULTS AND DISCUSSION**

The measurements of magnetization of the as-grown Py-films in this study exhibit two magnetic axes within the film plane with different coercitive field (H parallel to the easy and the hard magnetic axis). Such uniaxial inplane anisotropy is observed in those samples, which are prepared in a non vanishing magnetic field during preparation. All of these as-grown samples have the same easy axis perpendicular to the direction of substrate movement.

Figure 2 shows the typical field dependence of the normalized resistivity of an as-grown film with the external field parallel to the easy axis (e.a., curve(a)) and the hard axis (h.a., curve(b)), respectively. In both cases the sensing current I was applied perpendicular to the field (method (1) and (2) in Fig. 1). In the case of the field along the easy axis (curve(a)) the reversal of the magnetization with field occurs by nucleation of domains with moments antiparallel to the former direction together with the motion of 180°-walls. Because the AMR is proportional to $\cos^2\Theta$, with $\Theta$ as the angle between the magnetization and the current [11], there is no change in the AMR expected between antiparallel aligned moments. Therefore the resistivity should not change by applying a magnetic field along the easy axis. Experimentally we observe a small change (about 0.2%, curve(a)) which could be interpreted either by the field being not exactly aligned parallel to this direction or by a slight distribution of easy axes (see below). Along the magnetic hard axis the magnetic reversal occurs by coherent rotation of the magnetic moments with 90°-wall motion. In the hard axis direction the moments are perpendicular to the field direction for zero field. Thus increasing the field rotates the moments by 90° and one obtains a maximum of the AMR. According to curve(b) in Fig. 2 we measure a 5% change in the AMR. Applying the current I parallel to the magnetic field results in an equal change of the resistivity for both axes although the resistivity itself increases with field.

Figure 3 shows the resistivity measurements of the same sample as in Fig. 2 after annealing at 720K for 15 minutes. The field was aligned along the former easy axis with the sensing current I applied both parallel and perpendicular to H (curves(a) and (c)), respectively. The inset shows the corresponding measurements with the field along the former hard axis (curves(b) and (d)). After this heat treatment significant changes in $\rho/\rho_c$ along both the former easy as well as the former hard axis are observed compared to the same measuring geometry in the unannealed sample (see curve(a) and (b) in both Fig. 2 and Fig. 3). Now $\Delta\rho/\rho_c$ is even larger along the former easy axis. The fact that $\Delta\rho/\rho_c$ does not vanish completely along the former hard axis means that the uniaxial anisotropy of the as-prepared films has changed to a more isotropic behavior. Nevertheless, there is some dominance of the former hard axis after the anneal above 520K.

Figure 4 summarizes the evolution of $\rho_c$ and $\Delta\rho/\rho_c$ as a function of the annealing temperature $T_a$. The resistivity $\rho_c$ (Fig. 4a) decreases with $T_a$ up to 570K, stays about constant for higher temperatures, and increases again for $T_a > 920K$. In Fig. 4b the most prominent fact is a crossover of $\Delta\rho/\rho_c$ measured along the magnetic easy and hard axes at $T_a \approx 550K$ before both branches merge for temperatures around 870K. Along the easy direction there is hardly any AMR up to about 520K (see also Fig. 2, curve(a)). Above this temperature $\Delta\rho/\rho_c$ increases sharply and reaches even higher values than for as-deposited samples measured along the hard axis. On the other hand, the latter decreases above 520K but does not totally vanish. The crossover is also observed in the development of the sensitivity $(\Delta\rho/\rho_S)$/FWHM (not shown here), which shows qualitatively the same behavior as $\Delta\rho/\rho_c$ [12]. The maximum sensitivity observed in the range of 700K is 1.5%/Oe.

For a structural film characterization we performed small and wide angle X-ray diffraction measurements The as-sputtered samples mainly exhibit a <111> fiber texture with a random orientation of the crystallites within the plane. This was made sure by measurement of the

<311>-peak at an inclined angle, showing no dependence of peak position and peak intensity on the inplane orientation [12]. Thus the uniaxial anisotropy is not related to any observable preferential grain orientation within the plane of the film.

As already mentioned above the differences in resistivity measurements along the easy and the hard axis are well understood. The dependence of these directions on the orientation of the substrate relative to the direction of motion during the sputter process as well as on its position within the substrate holder (and thus the magnetic field strength and direction of the target gun) was verified by systematically changing both of these parameters, see [12]. We always obtain an uniaxial anisotropy for samples prepared on one half of the holder where they are exposed to the stray field of the target pointing in only one direction as a mean. For all these samples the easy axis was aligned perpendicular to the direction of motion during film deposition (and thus parallel to the mean stray field) and the hard axis was correspondingly aligned parallel to the direction of motion.

For an explanation of the above results we suggest that one origin for the experimentally observed anisotropy is related to anisotropic film stress generated during deposition. Bozorth [13] has already shown that stress can dramatically change the AMR. This is even true for one-component magnetic materials, e.g. Ni [14]. As a second effect, atomic chemical short range ordering can introduce an anisotropy (in this case at lower temperatures as compared to annealing in a magnetic field after preparation) which means that there exists a preferential orientation along one direction for alike-atom pairs (nearest neighbors), i.e. Fe-Fe and Ni-Ni-pairs [15,16]. As mentioned above, the field at the substrate is about $15 \times 10^{-4}$ T during film deposition which is sufficiently high to align the magnetic moments and to introduce an anisotropy of this kind for $Ni_{81}Fe_{19}$ [17]. However, in our sputtering device the direction of the stray field acting on the film is changing during the sputter process (by about 90 to 120 degrees depending on the exact position of the sample during preparation). Therefore the magnetically induced second anisotropy systematically changes direction during the whole film growth. Conse-quently this anisotropy is not simply uniaxial anymore. Thus the samples are magnetically inhomogeneous which might explain the large $\Delta\rho$. For a change of atomic pair ordering one has to overcome the rather high activation energy for atom diffusion ranging from 2.3eV to 2.7eV in $Ni_xFe_{1-x}$-alloys for x = 0.61 to 0.86 [18]. Therefore the observed change in anisotropy by the heat treatments at e.g. 570K for a few hours (Takahashi [9]) or respectively at 520K for 15min (our samples) is more likely to arise from a stress relaxation instead of a change of directional short range ordering. Only at higher temperatures (T > 720K) local reorientations are assumed to produce long range ordering ($L1_2$-structure) which corresponds to the merging effect as described in Fig. 4b. At this stage the differences measured along the different directions vanish.

## IV. CONCLUSIONS

$Ni_{81}Fe_{19}$-films, dc-magnetron sputtered in UHV, show a crossover behavior of magnetic anisotropy due to annealing at temperatures above 520K. This crossover is explained to arise from a change in the dominating source of anisotropy, assigned to mechanical stress and the preferred atomic pair ordering within the film. Both sources are generated by the magnetic stray field during film deposition but relax, however, at different temperature ranges.

## ACKNOWLEDGMENTS

We thank H.A.M. van den Berg for helpful discussions and U. Bete for technical assistance. This work was financially supported by the BMBF, Förderkennzeichen Nr. 13N6174.

**CAPTIONS OF FIGURES**

FIG. 1  Schematic view of the four measuring geometries with different orientations of applied magnetic field H, the measuring current I, and the easy and the hard axes, respectively, which are necessary to obtain the AMR and its orientation dependence within the film plane.

FIG. 2  Field dependence of the normalized resistivity $\rho/\rho_c$ (with $\rho_c$ taken at the coercive field $\mu_0 H_c$) of an as-prepared $Ni_{81}Fe_{19}$-film with the external field parallel to (a) the easy axis and (b) to the hard axis. In both cases the measuring current I is perpendicular to the magnetic field H.

FIG. 3  Field dependence of the normalized resistivity $\rho/\rho_c$ with the field parallel to the former easy axis and I parallel and perpendicular to the field, respectively, for the same sample as in Fig. 2 after an anneal at 720K for 15min. Inset: Measurements with H parallel to the former hard axis.

FIG. 4  Dependence of (a) the resistivity $\rho_c$ at the coercive field $\mu_0 H_c$ and (b) the normalized maximum change in resistivity $\Delta\rho/\rho_c$ of a 20 nm thick $Ni_{81}Fe_{19}$-film on the annealing temperature $T_a$. Annealing took place at the respective temperatures for 15min each, the measurements were performed at room temperature.

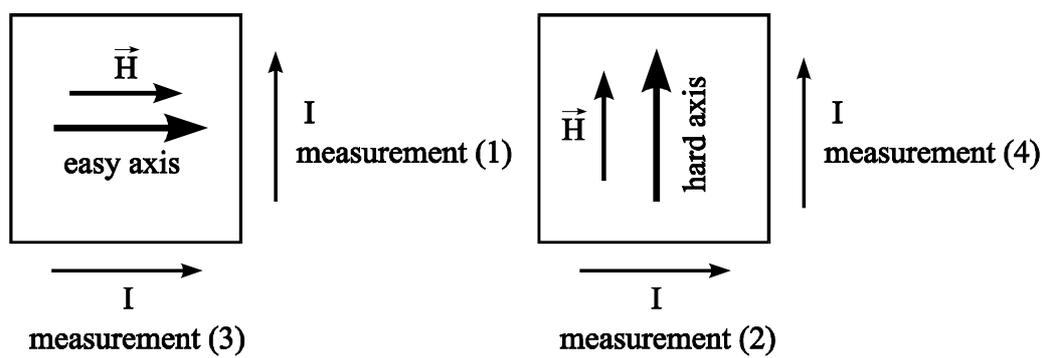

**FIG. 1**

T. Lorenz et al.

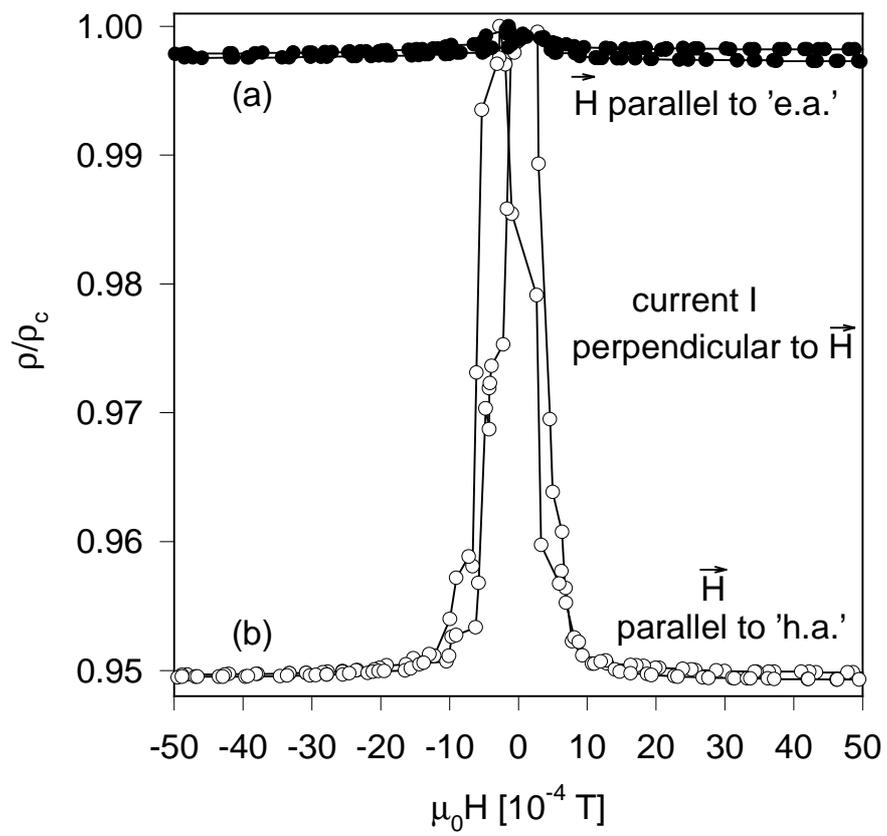

**FIG. 2**

T. Lorenz et al.

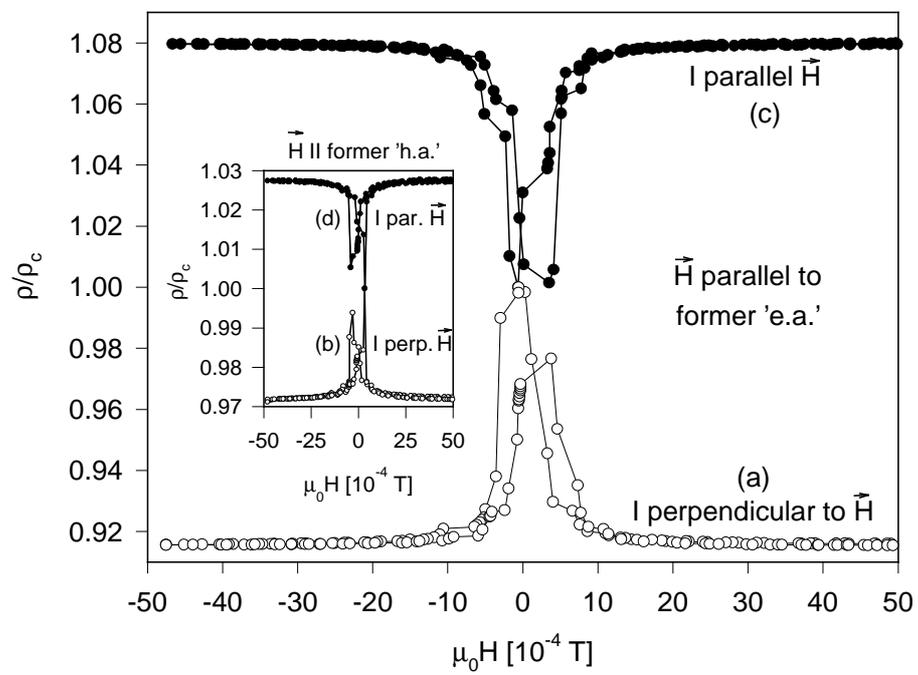

**FIG. 3**

T. Lorenz et al.

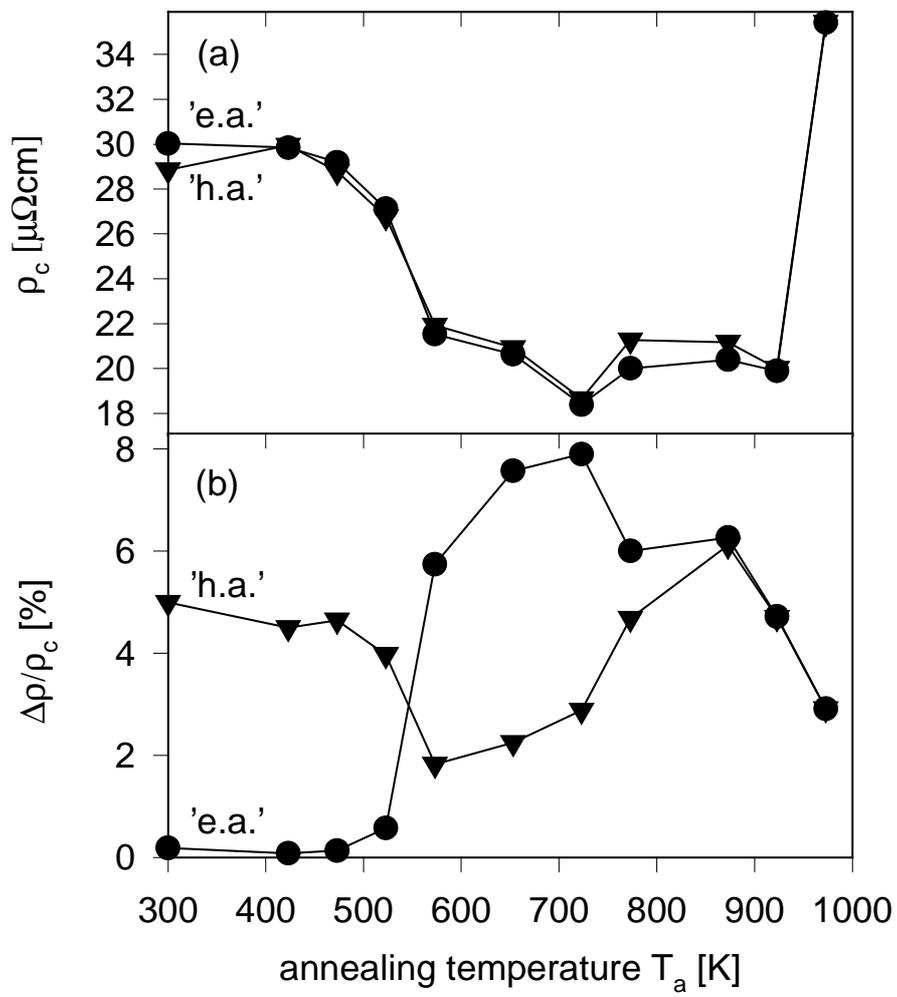

**FIG. 4**

T. Lorenz et al.